\documentclass[12pt]{article}
\usepackage{epsfig,latexsym}

\def\be{\begin{equation}}
\def\ee{\end{equation}}
\def\bea{\begin{eqnarray}}
\def\eea{\end{eqnarray}}

\begin{document}
\font\cmss=cmss10 \font\cmsss=cmss10 at 7pt

\hfill IC/2003/37 \vskip .1in \hfill CPHT-RR-027-0603 \vskip .1in \hfill hep-th/0306102

\hfill
\vspace{18pt}
\begin{center}
{\Large \textbf{On zero-point energy, stability and Hagedorn behavior of Type IIB strings on pp-waves}}
\end{center}

\vspace{6pt}

\begin{center}
{\textsl{F. Bigazzi $^{a}$, A. L. Cotrone $^{b,c}$}} 

\vspace{20pt}

\textit{$^a$ The Abdus Salam ICTP, Strada Costiera, 11; I-34014 Trieste, Italy.}\\
\textit{$^b$ Centre de Physique Th\'eorique, \`Ecole Polytechnique, 48 Route de Saclay; F-91128 Palaiseau Cedex, France.}\\
\textit{$^c$ INFN, Piazza dei Caprettari, 70; I-00186  Roma, Italy.}
\end{center}

\vspace{12pt}

\begin{center}
\textbf{Abstract }
\end{center}

\vspace{4pt} {\small \noindent
Type IIB strings on many pp-wave backgrounds, supported either by 5-form or 3-form fluxes, have negative light-cone zero-point energy. This raises the question of their stability and poses possible problems in the definition of their thermodynamic properties. After having pointed out the correct way of calculating the zero-point energy, an issue not fully discussed in literature, we show that these Type IIB strings are classically stable and have well defined thermal properties, exhibiting a Hagedorn behavior. 
}

\vfill
\vskip 5.mm
 \hrule width 5.cm
\vskip 2.mm
{\small
\noindent
bigazzif@ictp.trieste.it\\
Cotrone@cpht.polytechnique.fr}
%%%%%%%%%%%%%%%%%%%%%%%%%%%%%%%%%%%%%%%%%%%%%%%%%%%%%%%%%%%%%%%%%%%%%%%%%%%%%%%%%%%%
\section{Introduction}
Strings on pp-waves have been extensively studied in the last two years.
The importance of the maximally supersymmetric pp-wave resides in the fact that it arises as a Penrose limit of $AdS_5 \times S^5$ \cite{blaufig} and that IIB string theory on this background is exactly solvable \cite{M}.
This allowed to extend the tests of the $AdS/CFT$ correspondence, previously limited mainly to the supergravity approximation, to the full perturbative string spectrum on the pp-wave \cite{BMN}.
A lot of works has followed, extending this ``new'' correspondence to other known field theory/supergravity duals.
The Penrose limit always results in plane-waves, preserving at least 16 supersymmetries.
These are the so-called ``dynamical'' supercharges, which give no linearly realized supersymmetry on the worldsheet \cite{cve}.
If no other supersymmetry is preserved by the background, the string theory has a non trivial light-cone zero-point energy.
The aim of the present paper is to study the stability and the thermodynamics of strings with {\it negative} light-cone zero-point energy.
This is a common feature in a wide class of pp-wave backgrounds, including ones with RR 3-forms turned on \cite{all}.
Though our considerations apply in general, we will focus here on the simplest model, by considering a family of homogeneous plane wave backgrounds with null RR five form field strength. This is the class considered in \cite{blaufig} including the maximally supersymmetric pp-wave as a particular case
\begin{eqnarray}
ds^2 &=& 2dx^+dx^- - \mu^2\sum_{i=1}^8(a_i^2x_i^2)dx^+dx^+ + \sum_{i=1}^8dx^idx^i, \nonumber \\
&&F_{+1234}= F_{+5678}= 2\mu, \qquad  \sum_{i=1}^8a_i^2=8,
\label{pp}
\end{eqnarray}
the last condition on the constants $a_i$ being a consequence of the IIB supergravity equations of motion.
These backgrounds generically preserve 16 supersymmetries. When $a_i=1,i=1...8$, we recover the maximally supersymmetric pp-wave background with 32 Killing spinors. 
The superstring zero-point light-cone energy $E_0$ is generically not zero and varies as a function of the worldsheet mass parameter $m=\mu\alpha'p^+$.  

The first, apparently naive, problem we face concerns the way of calculating $E_0$. Some confusion can in fact arise when trying to apply the prescriptions available in literature to our general case. In all the papers on superstring partition functions on pp-waves at zero or finite temperature appeared till now only one method is considered. It amounts on treating bosonic and fermionic contributions separately and on renormalizing them. This means that for every periodic worldsheet field of mass $m$, for example,  the zero-point energy is evaluated in terms of a Casimir energy, i.e. by the following subtraction procedure
\be
\sum_{n=-\infty}^{\infty}\sqrt{n^2+m^2}\rightarrow \sum_{n=-\infty}^{\infty}\sqrt{n^2+m^2}-\int_{-\infty}^{\infty}dx\sqrt{x^2+m^2}.
\label{cas}
\ee  
As we will show, however, a brute application of this renormalization prescription amounts on subtracting a {\it finite} term to the whole $E_0$, which is {\it finite}, and this, as far as we understand, is not justified. This finite term has in principle a physical meaning, it contributes to the light-cone energy spectrum, to the superstring partition function and it is relevant when considering string/field theory duality, so it must be taken into account.

Our strategy, and our very general proposal, will be first to {\it evaluate} physical quantities as the whole zero-point energy or partition function, without referring to any renormalization scheme: if physical quantities are finite, no renormalization prescription is then needed.

As we said, we are interested in the cases when the superstring zero-point energy $E_0$ is negative. 
In the limit $\mu\alpha'p^+ <<1$ (this is usually referred to as the ``supergravity limit'' as the leading order contribution to the string motion comes from the zero-modes) we can study the consequences (or the meaning) of the negativity of $E_0$ by referring to supergravity. So, the question is: does the negative value of the zero-point energy correspond to some (classical) instability of the backgrounds here considered?  
We will see that in all the cases in which $E_0$ stays finite in the supergravity limit, no instability can be read from the supergravity perspective, no matter the sign of $E_0$. The situation is problematic only when $E_0$ goes to minus infinity for $\mu\alpha'p^+\rightarrow 0$ (just as in the Type 0B case considered in \cite{noi0}) where a tachyonic instability appears.

In the opposite regime $\mu\alpha'p^+>>1$ the contribution to $E_0$ coming from integrals as in (\ref{cas}) is the leading one. Our prescription for calculating zero-point energies should be thus relevant when studying string/field theory dualities (the regime  $\mu\alpha'p^+>>1$ usually corresponds to a weak effective coupling regime in the dual gauge theory sector) both at zero and at finite temperature. 
Leaving aside the interpretation in the possible dual gauge theory, we can explore the consequences of the zero-point energy behavior on some stringy quantities, like the partition function and free energy at finite temperature. Here, just as in the maximally supersymmetric context, we find indications that there is a Hagedorn temperature $T_H$. As usual, $T_H^{-2}$ is proportional to the string zero-point energy evaluated with fermions obeying antiperiodic boundary conditions. The finite term we take in $E_0$ modifies $T_H$ with respect to the result we would have obtained by applying the renormalization prescription (\ref{cas}).

The paper is organized as follows. In Section 2 we give the general prescription for calculating zero-point energies. Section 3 is devoted to the stability issues, and Section 4 to the thermodynamics of IIB strings on the background (\ref{pp}).
%%%%%%%%%%%%%%%%%%%%%%%%%%%%%%%%%%%%%%%%%%%%%%%%%%%%%%%%%%%%%%%%%%
\section{Zero-point energy for superstrings on pp-waves}

Let us consider Type IIB superstring on the backgrounds (\ref{pp}). In light-cone gauge
we find that, although 16 supersymmetries are preserved, these are the dynamical ones, so that there is no linearly realized worldsheet supersymmetry (except for the $a_i=1$ case). On the worldsheet in fact, the 8 fermions have all mass $m=\mu\alpha'p^+$ just as in the maximally supersymmetric case (fermionic masses only depend on the five-form field strength), while bosons have masses $m_i=a_im$\footnote{The quantization of the theory on the backgrounds (\ref{pp}) is identical to the one in the maximally supersymmetric case \cite{M,MT}, with the only differences of the worldsheet boson masses and the existence of a zero-point energy.}.

The IIB light-cone zero-point energy reads
\be
E_0={\mu\over2m}\sum_{n=-\infty}^{\infty}[\sum_{i=1}^8\sqrt{n^2+a_i^2m^2} -8\sqrt{n^2+m^2}].
\label{zeropoint}
\ee
The question we want to address here concerns the way of evaluating $E_0$. We find it, as an application of the Abel-Plana formula, in the sometimes referred to as the {\it Epstein formula}, which amounts on taking the $z=-1/2$ value of the expression
\begin{eqnarray}
F[z,s; m^2]&=& \sum_{n=1}^{\infty}\left[(n+s)^2 + m^2\right]^{-z}= {1\over2}[(s+1)^2 +m^2]^{-z} +
\int_1^{\infty}dx[(x+s)^2+m^2]^{-z} + \nonumber \\
&+& i\int_0^{\infty}dt\left[{ [(1+it+s)^2+m^2]^{-z}-
[(1-it+s)^2+m^2]^{-z}}\over{e^{2\pi t}-1}\right].
\label{epst}
\end{eqnarray}
 This is to be used when calculating the zero-point energy for a worldsheet field of mass $m$ and periodicity $\phi(t,\sigma+2\pi)=e^{2\pi is}\phi(t, \sigma)$. 

The Epstein formula is only an instrument for calculating\footnote{Formula (\ref{epst}) for $z\neq -1/2$ {\it regularizes} the $z= -1/2$ series and allows summing the bosonic and fermionic contributions in (\ref{zeropoint}). It is {\it not}, however, a {renormalization} prescription.} $E_0$. It includes (for $z=-1/2$) an evidently divergent term, the integral over $x$, and so it does not correspond to any renormalization procedure\footnote{This formula was used in \cite{noi0} to study the twisted sector zero-point energy for 0B strings in the $a_i=1$ pp-wave background.}. In the present context it gives
\be
E_0= {\mu\over2}\left[\sum_{i=1}^8|a_i|-8\right] +{\mu\over m}\left[\sum_{i=1}^8F[-1/2, 0; a^2_im^2]-8F[-1/2, 0; m^2]\right].
\label{e0}
\ee
Now, the point we would like to stress is that this expression is {\it finite}: though the individual contributions to the zero-point energy are infinite, all the divergences naturally disappear when considering the {\it whole} physical quantity. This is a generic feature for supersymmetric theories. The presence of both bosonic and fermionic fields guarantees $E_0$ to be finite without any need of subtracting divergences artificially. 

If we study the expression in (\ref{e0}) as a function of $m$ we find that $E_0$ is generically {\it negative} and it is zero only in the maximally supersymmetric case. 
In the small $m$ limit we find that the leading order contribution comes from the zero-modes and
\be
E_0\rightarrow {\mu\over2}(\sum_{i=1}^8|a_i|-8)\leq0,
\label{smalm}
\ee
the equality holding only for $a_i=1$.

In the large $m$ limit, as suggested by the Epstein formula (\ref{epst}), we can approximate the series in (\ref{zeropoint}) by an integral\footnote{This observation was used, for example, in \cite{froltse} in the context of semi-classical quantization of spinning superstrings.}
\be
E_0\approx{\mu\over 2m}\int_{-\infty}^{\infty}dx[\sum_{i=1}^8\sqrt{x^2+a_i^2m^2}-8\sqrt{x^2+m^2}]= -{\mu m\over2}\sum_{i=1}^8a_i^2\log|a_i|\leq0.
\label{largem}
\ee
This expression is {\it finite} for every finite value of $m$. It is zero only in the maximally supersymmetric case $a_i=1$. It is also {\it negative}, just as in the small $m$ regime. When $m\rightarrow\infty$, $E_0\rightarrow -\infty$.

The problem we face is that in literature we found another way of dealing with  zero-point energies for strings on pp-waves. It is based on the renormalization prescription appearing in \cite{gbg} and in a lot of papers which followed (sometimes we will refer to it as the $\Delta$ prescription). This amounts on renormalizing (a la Casimir) each single contribution to $E_0$ as
\begin{eqnarray}
&&{1\over2}\sum_{n=-\infty}^{\infty}\sqrt{(n+s)^2 + m^2}\rightarrow \Delta(s ; m)=-{1\over 2\pi^2}\sum_{p=1}^{\infty}\int_0^{\infty}dy\, e^{-yp^2-{\pi^2m^2\over y}}\cos(2\pi ps)\nonumber \\
&&={1\over2}\sum_{n=-\infty}^{\infty}\sqrt{(n+s)^2 + m^2}-{1\over2}\int_{-\infty}^{\infty}dx\sqrt{(x+s)^2 + m^2}.
\label{delta}
\end{eqnarray}
Once each contribution is $\Delta$-renormalized\footnote{In the massless case this gives $\Delta(s; 0)= 1/24 - (1/8)(2s-1)^2$ from which the values $-1/12, 1/24$ for periodic and anti-periodic fields follow.}, the whole zero-point energy is evaluated. Generically this gives a different result with respect to the one   obtained above, as now, by using (\ref{delta}) we get
\be
E_0 = {\mu\over m}\left[\sum_{i=1}^8\Delta(0; a_im) - 8\Delta(0; m)\right].
\ee
This expression too is generically negative.
In the small $m$ limit we find the same leading order behavior as in (\ref{smalm}), but in the large $m$ limit we get a very different result with respect to the one found by using the Epstein calculation tools. In fact it is evident that in the limit $m\rightarrow\infty$ $\Delta(m)$ goes to zero by construction.
Thus, having used this method to calculate the zero-point energy, we would have found 
\be
E_0\rightarrow0 \qquad \rm{for} \qquad m\rightarrow\infty,
\ee
i.e. a result which evidently contradicts the previous one.

The contradiction disappears in the maximally supersymmetric case $a_i=1$ (and, for more generic backgrounds, whenever ${\cal N}>16$ supersymmetries are preserved) where both methods give $E_0=0$ identically, as bosons and fermions have both the same mass $m$. This is in fact the case considered in \cite{gbg}. Though in that paper the $\Delta$ renormalization method is not given as a general prescription for calculating zero-point energies, we found the point a bit confusing when trying to study more general situations.

As the reader can see, the $\Delta$ renormalization procedure is subtracting a specific {\it finite} term (the integral in (\ref{largem})) to the {\it finite} superstring zero-point energy. However in superstring theory one must not use ad hoc renormalization prescriptions -- and none is needed -- as long as the sum in $E_0$ is finite. One just need to estimate it, and the Epstein method is a way of doing that\footnote{We thank A. Tseytlin for an useful discussion on this point.}. 

We thus suggest that this method (or analogous ones which do not subtract anything from the superstring zero-point energy) is the correct one. This should be taken into account when considering two kinds of problem:
\begin{itemize}
\item{ String/gauge theory correspondence. As we know from the standard BMN case, the parameter $m$ is generically connected to the effective coupling $\lambda'$ of the gauge theory sector which is conjectured to be dual to the string-on-pp-wave model. In all the examples studied in literature $\lambda'\rightarrow0$ when $m\rightarrow\infty$. Thus our previous observation is relevant when trying to compare string theory results with weak coupling field theory ones.}
\item{Strings on pp-wave backgrounds at finite temperature. In this case, for example, the value of the zero-point energy is important when calculating the Hagedorn temperature or free energy. Thus the behavior of, say, $T_{H}$, is surely affected by the way of calculating $E_0$.}
\end{itemize}
%%%%%%%%%%%%%%%%%%%%%%%%%%%%%%%%%%%%%     
\section{Negative energy versus classical (in)stability}
As we have commented, apart from the maximally supersymmetric case, we find that the superstring ground state has negative light-cone energy. What does this mean? Is it connected to some classical instability of the pp-wave background we are considering? The question is sensible, because we find $E_0<0$ also in the small $m$ limit, i.e. when supergravity approximation holds. 

The backgrounds we are considering preserve 16 supersymmetries and thus one would be tempted to conclude that they have to be stable. But, as observed in \cite{safin}, due to the fact that pp-wave backgrounds are not asymptotically flat, we cannot trivially extend here the standard arguments relating supersymmetry and stability. 
In flat space the light-cone energy gives the space-time mass spectrum for the string states and a negative value signals the presence of a 10-D tachyon (see for example the Type 0 case). In a pp-wave background the situation, in principle, is different.

We will ask here whether the excitation of a field in the supergravity sector of string theory is stable or not. This will be done by considering the expression of the light-cone energy $p_+$ as calculated in string theory, and then calculating the expected light-cone energy of the corresponding supergravity field.
From the comparison of the two, we will be able to cast conclusions on the (in)stability of this mode.
For example, the light-cone energy of the twisted sector ground state of Type 0B string theory in the small $m$ limit is of the form $p_+=4\mu -1/\alpha'p_-$ \cite{noi0}.
The field having this energy is a scalar, and from the study of its supergravity equations of motion (e.o.m.) one realizes that the energy above is the one for a $massive$ scalar on the pp-wave, with $negative$ squared mass $-2/\alpha'$.
As it will turn out, a scalar with negative squared mass is an unstable mode on the pp-wave, so we conclude that the twisted sector of Type 0B theory is unstable.

As a stability criterion\footnote{This is a commonly used criterion for classical stability, see for example \cite{mtaylor}.}, we will ask whether a supergravity mode has or has not an exponentially increasing behavior with time. This was adopted in a pp-wave context by \cite{safin,pando1} to study the stability of plane-wave backgrounds with no flux turned on; these are potentially problematic because the supergravity e.o.m. require $\sum_{i=1}^8a_i^2=0$, which means that some of the $a_i$ has to be imaginary. This in turn implies the presence of negative mass squared worldsheet fields, and rises the question of stability in a natural way.
Nevertheless, in \cite{safin,pando1} it was argued that those backgrounds are classically stable. 
In our case, with a five-form turned on, we can solve the supergravity e.o.m. with all $a_i$ being real, and so we can avoid worldsheet ``tachyons'' in light-cone gauge. This is the case we will consider here, taking in mind that a more general solution can involve negative squared mass, and referring to \cite{safin,pando1} for an analysis of this case. 
Thus, the only problem we face here is the negativity of $E_0$. 

As a clarifying example, let's see what happens in flat space-time for a scalar excitation $\phi$ of mass $m$. The light-cone dispersion relation reads $p_+p_-=m^2$. Now, for this mode to be stable, both $p_+$ and $p_-$ must be real, otherwise $\phi$ would explode in some direction (i.e. it would not be $\delta$ normalizable). At first sight, thus, there is nothing wrong here if $m^2<0$, but this is not the case of course, the light-cone coordinates somewhat hiding the inconsistency. If we go to Cartesian coordinates $z,t={1\over \sqrt{2}}(x^+ \pm x^-)$, the dispersion relation reads $\omega=\pm\sqrt{p_z^2+m^2}$, so that a stable mode has always $m^2\geq 0$, of course.

So let's move to the case of interest (\ref{pp}) and start from a scalar excitation again.
The equation of motion is
\begin{equation}
(\Box-m^2)\phi \equiv \left[2\partial_+ \partial_- +
\mu^2(\sum_{i=1}^8a_i^2x_i^2)\partial_- \partial_- + \partial_i\partial_i -m^2\right]\phi = 0. \label{eqmot}
\end{equation}
We are interested in those solutions which go to zero when $x^i\rightarrow\infty$. 
The simplest of them (see also \cite{Leigh}) is\footnote{The general
normalizable solution contains a product of Hermite polynomials depending on
the transverse coordinates. The choice (\ref{cho}) corresponds
to the lowest of such polynomials; it is the same as taking the transverse momenta $p_i=0$ in the flat case. We consider only this excitation because we are interested 
in the string theory
vacuum, without any insertion of bosonic zero-mode
oscillators. } 
\begin{equation} \label{cho}
\phi\approx e^{i(p_+x^+ -p_-x^-)}e^{-{\mu\over2}\sum_{i=1}^8|p_-a_i|x_i^2}.
\end{equation}
Its light-cone energy reads 
\begin{equation}\label{tachi}
p_+={\mu\over2}\sum_{i=1}^8|a_i| + m^2/2p_-,
\end{equation}
i.e. it is the (semi-)sum of the bosonic zero-modes, with the mass correction.
Again, it seems that a negative squared mass would be allowed. 
But, again, this is misleading.

In Cartesian coordinates 
the solution reads
\begin{equation}
\phi\approx e^{i(\omega t -p_zz)}e^{{\mu\over 2\sqrt{2}}(\omega-p_z)(\sum_{i=1}^8|a_i|x_i^2)},
\end{equation}
which has real $\omega$ if
\begin{equation}
p_z^2-{\mu\sum_{i=1}^8|a_i|\over \sqrt{2}}p_z+{\mu^2\over8}(\sum_{i=1}^8|a_i|)^2+m^2 \geq 0.
\end{equation}
This is true for every real $p_z$ only for $m^2 \geq 0$.
As we pointed out above, the ground state of the twisted sector in Type 0B string theory is a scalar with energy behaving as in (\ref{tachi}) in the supergravity limit, with $negative$ squared mass, so that it is a real unstable mode, i.e. a tachyon \cite{noi0}.

The ground state in the Type IIB string on the backgrounds (\ref{pp}) is not a scalar, but a combination of the trace of the $SO(4)$ part of the graviton with the (pseudo)scalar part of the five-form, call it $h$ \cite{MT}.
This excitation satisfies a slightly modified equation w.r.t. the purely scalar one, having as a correction a $8i\mu\partial_-$ piece coming from the five-form coupling\footnote{One can check that the e.o.m. for this mode on the background (\ref{pp}) is the same as the one derived in \cite{MT}.}.
Then the solution of the e.o.m. is still in the form (\ref{cho}) but now with a light-cone energy (we take $m^2=0$)
\begin{equation}
p_+={\mu\over2}\left(\sum_{i=1}^8|a_i|-8\right).
\end{equation}
This is the (semi)difference of the bosonic and fermionic zero-modes, and it is exactly the supergravity limit of the string zero-point energy, eq. (\ref{smalm}).
In general it is $negative$.
But the important point is that in Cartesian coordinates the dispersion relation is
\begin{equation}
\omega=p_z+{\mu\over\sqrt{2}}\left(\sum_{i=1}^8|a_i| -8\right),
\end{equation}
giving a real $\omega$ for every real $p_z$.
Then the string theory is classically stable even if it has a negative light-cone zero-point energy. 

We can conclude that whenever the string zero-point light-cone energy in the supergravity limit goes to a constant independent on $p_-$, the field corresponding to the string ground state is massless. The Cartesian dispersion
relation associated to it is thus always solvable in terms of real parameters, and so no classical instability appears, no matter the sign of $E_0$.
In the case in which $E_0\rightarrow const + k/2p_-$ (this is the case in the twisted sector of Type 0B string on pp-wave) the field corresponding
to the string ground state is massive with $m^2=k$. This is an unstable mode if $k<0$.

These considerations are very general and so also apply to a lot of physically relevant pp-wave backgrounds with also 3-form fluxes turned on \cite{all}. We will explore in detail some of these cases in a forthcoming paper \cite{abc}.

\section{Hagedorn behavior of strings on homogeneous pp-waves}
The same arguments used in \cite{Hammou} let us conclude that the one-loop partition function for IIB strings on the background (\ref{pp}) is zero. This is immediately recognizable in the path integral formulation, where the contributions coming from the mass terms are integrated out so that the one-loop partition function equals the flat space-time one. An equivalent result can be obtained in the operatorial formalism \cite{Takayanagi,Hammou}. Also, the modular invariance is immediately guaranteed. Since the zero-temperature case is trivial, let us see what happens at finite temperature.

Following the grand canonical conventions of \cite{greene}, 
we can construct a globally time-like Killing vector for the background (\ref{pp})
\be
\xi(a, b)= a\partial_+ + b\partial_- , \qquad  a,b>0
\ee
and thus formally write the string partition function as
\be
{\cal Z}(a, b; \mu) = Tr e^{-a p_-  -b p_+}.
\ee
The dimensional parameters $a,b$ are related to the inverse temperature $\beta=T^{-1}$ and the chemical potential $\nu$ by \cite{brower}
\be
\beta = {{a+b}\over\sqrt2}, \qquad \nu = {{b-a}\over\sqrt2}.
\ee
By referring to \cite{greene,brower} for all the details of the calculation, we find that the free energy $F$ for IIB strings on the generalized pp-wave background (\ref{pp}) is given by $F=-T\log{\cal Z}$, where
\be
\log{\cal Z}= -{a\over2\pi\alpha'}\int d\lambda dm\int_{-1/2}^{1/2}d\tau_1\int_{0}^{\infty}{d\tau_2\over\tau_2^2}{\Theta^4_{1/2,0}(\tau,{\bar\tau};m)\over\prod_{i=1}^8\Theta^{1/2}_{0,0}(\tau,{\bar\tau};a_im)}\sum_{r=odd}^{\infty}e^{-{ab r^2\over2\pi\alpha'\tau_2} +2\pi i\lambda(m-\mu a r/2\pi\tau_2)} .
\label{free}
\ee
Here
\be
\Theta_{k,s}(\tau,{\bar\tau};m)= e^{4\pi\tau_2E(s;m)}\prod_{n=-\infty}^{\infty}(1-e^{-2\pi\tau_2|\omega_{n+s}|+2\pi\tau_1(n+s)+2\pi ik})(1-e^{-2\pi\tau_2|\omega_{n-s}|+2\pi\tau_1(n-s)-2\pi ik}),
\ee
where $E(s;m)$ is (proportional to) the zero-point energy for a worldsheet field of mass $m$ and periodicity set by $s$ as usual. Also, $\omega_{n+s}=\sqrt{(n+s)^2+m^2}$. 

In literature $E(s; m)$ is calculated by using the $\Delta$ prescription, thus avoiding divergent contributions to the functions $\Theta$. But what we claim here is that the important object to look at is {\it the whole superstring partition function}, and this receives a {\it finite} contribution from the integral terms subtracted by the $\Delta$ prescription.
So, let us take
\be
E(s; m)= \Delta(s ;m)+{1\over2} \int_{-\infty}^{\infty}dx\sqrt{(x+s)^2+m^2}=\Delta(s;m)+ {1\over2}\int_{-\infty}^{\infty}dx\sqrt{x^2+m^2} .
\label{enz}
 \ee
The reader can easily check that the integral term in (\ref{enz}) does not change the transformation properties of the $\Theta$ functions under modular transformations and thus does not spoil the modular invariance of the theory\footnote{By referring to the notations of \cite{gbg}, the computational method we adopt to calculate zero-point energies only modifies the $f$-functions as $f_i(m;q)\rightarrow q^{-I(m)}f_i(m;q), \quad i=1,2,3,4$ where $2I(m)= \int_{-\infty}^{+\infty} dx \sqrt{x^2 + m^2} = \int_{-\infty}^{+\infty} dx \sqrt{(x+1/2)^2+ m^2}$. 
 This term is the same for periodic and anti-periodic fields. Thus
 it does not change the whole IIB partition function in the maximally
 supersymmetric case. It also does not break any modular invariance, as 
 $q^{-I(m)} = e^{2\pi t I(m)} = e^{2\pi I(tm)/t}$. 
 This way, formulae (3.21) of \cite{gbg} for the modular transformations
 of the $f_i$ , and so of the $\Theta$ functions, are still exactly respected.}.
 
Now, let us study the $\tau_2\rightarrow0$ limit of (\ref{free}) while taking
$m\rightarrow\infty$ in order to have $M=m\tau_2$ fixed. We also keep fixed the ratio $\theta\equiv\tau_1/\tau_2$.
Following the same steps as in \cite{greene} we find that
\be
{\Theta^4_{1/2,0}(\tau,{\bar\tau};m)\over\prod_{i=1}^8\Theta^{1/2}_{0,0}(\tau,{\bar\tau};a_im)}\approx e^{2\pi\tau_2I_{tot}(M)}e^{-{8M\over|\tau|}f(M,\theta,1/2)}e^{\sum_{i=1}^8{a_iM\over|\tau|}f(a_iM,\theta,0)},
\label{ta}
\ee
where
\be
I_{tot}(M)= {1\over2}\int_{-\infty}^{\infty}dx\left[8\sqrt{x^2+M^2}-\sum_{i=1}^8\sqrt{x^2+a_i^2M^2}\right]
\label{et}
\ee
and
\be
f(M;a)\equiv -{2\pi\over M}\Delta(a;M) .
\label{fun}
\ee
Let us stress that in the large $m$ limit the first term in (\ref{enz}) goes to zero exponentially, and thus only the second one contributes to (\ref{ta}). Note that having used the standard recipe with only the $\Delta$ functions appearing in $E$ we would have had zero instead of $I_{tot}$ in (\ref{ta}). Here, instead, we find a zero large $m$ result only in the maximally supersymmetric case (where $I_{tot}$ is automatically zero), thus finding perfect agreement with all the papers on the argument appeared till now.

Finally, the whole integrand in (\ref{free}) converges when $\tau_2\rightarrow0$ (the relevant contribution in this case coming form the $r=1$ term) if 
\be
2ab> 2a\mu\alpha'\left[\sum_{i=1}^8a_if({a_ia\mu\over2\pi},0)-8f({a\mu\over2\pi},1/2)\right]+\mu^2a^2\alpha'\sum_{i=1}^8a_i^2\log|a_i|,
\ee
the last term being the contribution coming from $I_{tot}$. We thus find an indication that strings on homogeneous pp-waves have a Hagedorn temperature implicitly given by 
\be
\beta_H^{2}= \nu^2 + 8\pi^2\alpha'E_{tot}[{\mu(\beta_H-\nu)\over2{\sqrt2}\pi}],
\label{th}
\ee
where we used (\ref{fun}) and defined $E_{tot}[m]=8E(1/2;m)-\sum_{i=1}^8E(0;a_im)$.
In the flat space-time limit $\mu=0$, at zero chemical potential,  we recover the standard IIB Hagedorn temperature $T_H^{-2}=8\pi^2\alpha'$.

It is not clear if $T_H$ should correspond to a limiting temperature or to a phase-changing point. For some considerations about this issue in the maximally supersymmetric case\footnote{For other studies of finite temperature strings on pp-waves see \cite{tutti}.} we refer to \cite{pandovaman,greene,brower,grignani} which however give different claims on the argument. For example, in \cite{greene} the free energy at $T_H$ was shown to diverge, suggesting that $T_H$ could be interpreted as a limiting temperature. In \cite{brower} a seemingly more detailed calculation showed that the free energy is finite at the Hagedorn temperature, a necessary (but not sufficient) condition for $T_H$ being associated to some phase transition.

%%%%%%%%%%%%%%%%%%%%%%%%%%%%%%%%%%%%%%%%%%%%%%%%%%%%%%%%%%%%%%%%%%%%%%%%%%%%%%%%%%%%%%%%%%%%%%%

%%%%%%%%%%%%%%%%%%%%%%%%%%%%%%%%%%%%%%%%%%%%%%%%%%%%%%%%%%%%%%%%%%%%%%%%%%%%%%%%%%%%%%%%%%%%%%%

\begin{center}
{\large  {\bf Acknowledgments}}
\end{center}
We are grateful to A. Tseytlin for very useful comments and observations. We thank M. Blau, M. Gaberdiel, L. Girardello, M. B. Green, G. Grignani, A. Hammou, L. Martucci, L. A. Pando-Zayas, P. Silva, A. Zaffaroni for discussions. F. B. is partially supported by INFN.


\begin{thebibliography}{99}
\bibitem{blaufig}M. Blau, J. Figueroa-O'Farril, C. Hull, G. Papadopoulos, {\it A new maximally supersymmetric background of IIB superstring theory}, JHEP 0201 (2002) 047; hep-th/0110242. {\it Penrose limits and maximal supersymmetry},  hep-th/0201081.
\bibitem{M} R. R. Metsaev, {\it Type IIB Green-Schwarz superstring in plane wave Ramond-Ramond background}, Nucl.Phys. B625 (2002) 70-96; hep-th/0112044. 
\bibitem{BMN}D. Berenstein, J. Maldacena, H. Nastase, {\it Strings in flat space and pp waves from ${\cal N}=4$ Super Yang Mills}, JHEP 0204 (2002) 013; hep-th/0202021.
\bibitem{cve}M. Cvetic, H. Lu, C.N. Pope, {\it Penrose Limits, PP-Waves and Deformed M2-branes}, hep-th/0203082.
\bibitem{all}N. Itzhaki, I. R. Klebanov, S. Mukhi, {\it PP wave limit and enhanced supersymmetry in gauge theory}, JHEP 0203 (2002) 048; hep-th/0202153.\\
E. G. Gimon, L. A. Pando-Zayas, J. Sonnenschein, M. J. Strassler, {\it A Soluble String Theory of Hadrons}, hep-th/0212061.
\bibitem{noi0}F. Bigazzi, A. L. Cotrone, L. Girardello, A. Zaffaroni, {\it PP-wave and Non-supersymmetric Gauge Theory}, JHEP 0210 (2002) 030; hep-th/0205296. 
\bibitem{MT}R. R.  Metsaev, A. A. Tseytlin, {\it Exactly solvable model of superstring in plane wave Ramond-Ramond background}, Phys.Rev. D65 (2002) 126004; hep-th/0202109.  
\bibitem{froltse}S. Frolov, A. A. Tseytlin, {\it Semiclassical quantization of rotating superstring in $AdS_5 \times S^5$}, JHEP 0206 (2002) 007;  hep-th/0204226. {\it Multi-spin string solutions in $AdS_5\times S^5$}, hep-th/0304255.
\bibitem{gbg}O. Bergman, M. R. Gaberdiel, M. B. Green, {\it D-brane interactions in type IIB plane-wave background}, JHEP 0303 (2003) 002; hep-th/0205183. 
\bibitem{safin}D. Brecher, J. P. Gregory, P. M. Saffin, {\it String theory and the Classical Stability of Plane Waves}, Phys.Rev. D67 (2003) 045014; hep-th/0210308.
\bibitem{mtaylor}M. M. Taylor-Robinson, {\it Semi-classical stability of supergravity vacua}, Phys.Rev. D55 (1997) 4822-4838; hep-th/9609234.
\bibitem{pando1}D. Marolf, L. A. Pando Zayas, {\it On the Singularity Structure and Stability of Plane Waves}, JHEP 0301 (2003) 076; hep-th/0210309. 
\bibitem{Leigh}R. G. Leigh, K. Okuyama, M. Rozali, {\it PP-Waves and Holography},Phys.Rev. D66 (2002) 046004; hep-th/0204026.
\bibitem{abc}R. Apreda, F. Bigazzi, A. L. Cotrone, to appear.
\bibitem{Hammou} A. B. Hammou, {\it One Loop Partition Function in Plane Waves R-R Background}, JHEP 0211 (2002) 028; hep-th/0209265.
\bibitem{Takayanagi} T. Takayanagi, {\it Modular Invariance of Strings on PP-Waves with RR-flux}, JHEP 0212 (2002) 022; hep-th/0206010.
\bibitem{greene} B. R. Greene, K. Schalm, G. Shiu, {\it On the Hagedorn behaviour of pp-wave strings and ${\cal N}=4$ SYM theory at finite R-charge density}, Nucl.Phys. B652 (2003) 105-126; hep-th 0208163.
\bibitem{brower} R. C. Brower, D. A. Lowe, C.-I Tan, {\it Hagedorn transition for strings on pp-waves and tori with chemical potentials}, Nucl.Phys. B652 (2003) 127-141; hep-th/0211201.
\bibitem{tutti}Y. Sugawara, {\it Thermal amplitudes in DLCQ superstrings on pp-waves}, Nucl.Phys. B650 (2003) 75; hep-th/0209145.
Y. Sugawara, {\it Thermal partition function of superstring on compactified pp-wave}, hep-th/0301035.
S. Hyun, J.-D. Park, S.-H. Yi, {\it Thermodynamic behavior of IIA string theory on a pp-wave}, hep-th/0304239.
\bibitem{pandovaman}L. A. Pando Zayas, D. Vaman, {\it Strings in RR Plane Wave Background at Finite Temperature}, hep-th/0208066.
\bibitem{grignani}G. Grignani, M. Orselli, G. W. Semenoff, D. Trancanelli, {\it The superstring Hagedorn temperature in a pp-wave background}, hep-th/0301186.

\end{thebibliography}
\end{document}